\documentclass[aps,prb,preprint,groupedaddress]{revtex4}
\usepackage{graphicx}
\begin{document}

\title{Density of bulk trap states in organic semiconductor crystals: discrete levels induced by oxygen in rubrene}

\author{C. Krellner}
\altaffiliation{Present address: Max Planck Institute for Chemical Physics of Solids, D-01187 Dresden, Germany}
\author{S. Haas}
\author{C. Goldmann}
\altaffiliation{Present address: Philips Research Laboratories Aachen, D-52066 Aachen, Germany}
\author{K. P. Pernstich}
\author{D. J. Gundlach}
\altaffiliation{Present address: National Institute for Standards and Technology (NIST), Gaithersburg, MD, USA}
\author{B. Batlogg}
\email{batlogg@solid.phys.ethz.ch}
\affiliation{Laboratory for Solid State Physics, ETH Zurich, CH-8093
  Zurich, Switzerland}

\date{\today}

\begin{abstract}
The density of trap states in the bandgap of semiconducting organic single crystals has been measured quantitatively and with high energy resolution by means of the experimental method of temperature-dependent space-charge-limited-current spectroscopy (TDSCLC). This spectroscopy has been applied to study bulk rubrene single crystals, which are shown by this technique to be of high chemical and structural quality. A density of deep trap states as low as $\sim 10^{15}$\,cm$^{-3}$ is measured in the purest crystals, and the exponentially varying shallow trap density near the band edge could be identified (one decade in the density of states per $\sim$ 25\,meV). Furthermore, we have induced and spectroscopically identified an oxygen-related sharp hole bulk trap state at 0.27 eV above the valence band.
\end{abstract}

\maketitle

\section{Introduction}

The performance of organic thin-film transistors (OTFTs) is steadily improving, and the charge carrier mobility, as a key figure of merit, has reached values comparable to that of hydrogenated amorphous silicon. \cite{Horowitz1998, Gundlach02, Dimi01, Kelley2003, Klauk02} As with most semiconductors, the electrical performance is determined to a high degree by modifications of the ideal crystal, such as intentional doping and other chemical or structural effects which create electrically active states in the band gap. For a thorough understanding of the intrinsic capabilities and limitations of organic semiconductors, it is now highly desirable to quantitatively study such in-gap states, their density of states (DOS) spectrum, their origin, and their stability.

Field-effect transistors (FETs) on organic single crystals are well suited for determining the surface properties, with mobilities near 20\,cm$^2$/V s routinely achieved (for a review see Gershenson {\it et al.} \cite{Gershenson:2006}). Bulk properties, however, cannot be determined with this surface-sensitive method and thus one has to use alternative techniques, such as the time-of-flight method as shown in the pioneering work by Karl and co-workers, \cite{Warta1985} thermally stimulated current, or space-charge-limited-current (SCLC) measurements. For instance, SCLC measurements with coplanar electrodes were used by Lang {\it et al.} \cite{Lang01} to detect a metastable trap state in pentacene single crystals, where the SCLC changes over several orders of magnitude, indicating trap filling. 

Additional experimental techniques are used to detect defect states in organic semiconductors: Recently, oxygen-related states at the surface of naturally oxidized rubrene single crystals were detected by photoluminescene measurements. \cite{Mitrofanov2006} Kelvin-probe force microscopy not only offers a direct imaging of the potential across the channel of an OTFT, but additionally allows one to extract the DOS spectrum at the semiconductor/insulator interface. \cite{Tal2005}

In this study we go beyond the basic concept of SCLC and use the spectroscopic character of temperature-dependent SCLC spectroscopy (TDSCLC) as described in the theory by Schauer {\it et al.} \cite{Schauer04} to derive the bulk DOS spectrum in ultrapure rubrene single crystals. We found that these crystals of high structural and chemical quality have a broad distribution of deep states with a low total density of trap states, in addition to a steep band tail. We also use the spectroscopic technique to identify the energetic position of oxygen-induced bulk trap states in rubrene single crystals.

\section{Method}

A central aspect of TDSCLC is to exploit the spectroscopic character inherent in the temperature dependence of the SCLC due to the energy window associated with the Fermi-Dirac statistics. \cite{Schauer04} It is assumed that the SCLC is dominated by the charge carriers that are thermally excited from a localized trap into delocalized band states. Therefore the valence band edge is seen as separating localized and delocalized states, and it is chosen as the reference point for the energy scale ($E_{\rm v}=0$), with positive energy toward the midgap. The basic equations are Ohm's law in the form $j=e\mu_0n_{\rm f}(x)F(x)$ and the Poisson equation ${\rm d}F/{\rm d}x=-en_{\rm s}(x)/(\epsilon\epsilon_0)$. Here $F(x)$ is the electric field strength along the direction $x$ of current flow, $j$ is the current density, $\epsilon\epsilon_0$ is the product of the dielectric constant and electric permittivity (3.5 for rubrene), $\mu_0$ is the microscopic band mobility, $n_{\rm f}(x)$ is the density of free carriers, $n_{\rm s}(x)$ is the total density of carriers (free and trapped), which is given by the convolution of the density of trap states $h(E)$ in the energy gap with the Fermi-Dirac function $f(E,E_{\rm F},T)$, i.e., $n_{\rm s}=\int_E h(E)f(E,E_{\rm F},T)dE$. The shape of the current-voltage characteristic $j(U)$ reflects the increment of the space charge with respect to the shift of the Fermi energy and thus mirrors the energy dependence of the DOS,
\begin{equation}
\frac{dn_{\rm s}}{dE_{\rm F}} =\frac{1}{k_{\rm B}T}\frac{\epsilon\epsilon_0}{eL^2}\frac{(2m-1)}{m^2}(1+C)
\label{dns}
\end{equation}
with
\begin{equation}
C=\frac{B(2m-1)+B^2(3m-2)+d\bigl{[}\ln(1+B)\bigr{]}/d\ln U}{1+B(m-1)}{\rm .}
\label{C}
\end{equation}
Here $L$ is the thickness of the crystal with electrodes on opposite faces, $U$ the applied voltage, $k_{\rm B}$ the Boltzmann constant, $m=d\ln j/d\ln U$ the logarithmic slope of the $j(U)$ curve, and $B$ contains higher-order derivatives of $j(U)$, $B=-\bigl{[}dm/d\ln U\bigr{]}/\bigl{[}m(m-1)(2m-1)\bigr{]}$. The right hand side of Eq. (\ref{dns}) can be calculated from the current-voltage characteristic measured at only one temperature. For a complete reconstruction of the DOS, however, it is necessary to relate a given voltage $U$ to the energy of states which are being filled at this value of $U$. A first starting point is to extract an activation energy $E_{\rm A}$ from the Arrhenius plot of the TDSCLC data for a given $U$, i.e. $E_{\rm A}=-d\ln j/d(k_{\rm B}T)^{-1}$. Because $h(E)$ in general is not symmetric around $E_{\rm F}$, \cite{Arkhipov01} the energy $E_{\rm D}$ of the incremental change of space charge is slightly shifted from $E_{\rm A}$, typically by a fraction of $k_{\rm B}T$ \cite{Schauer03}
\begin{equation}
E_{\rm D}=E_{\rm A}+\frac{(3-4m)n}{(2m-1)(m-1)m}k_{\rm B}T{\rm .}
\label{Ed}
\end{equation}
Here $n\nolinebreak=\nolinebreak-d(E_{\rm A}/k_{\rm B}T)/ d\ln U$ is the derivative of the activation energy with respect to the applied voltage. To extract the DOS from the shape of the $j(U)$ curves, it is necessary to deconvolute Eq. (\ref{dns}) with respect to $df/ dE_{\rm F}$, using a high accuracy deconvolution method based on spline functions. \cite{Deutsch01}

\section{Experimental details}

The rubrene crystals were grown by physical vapor transport \cite{Laudise1998} under a stream of ultrapure Argon 6.0. The starting material (Aldrich purum) was sublimed three times in vacuum. Considerable effort was made to avoid contamination with any organic substances, e.g., by using glass tubes cleaned in acids. Typical crystals are platelike, where the direction perpendicular to the surface corresponds to the long axis of the orthorhombic unit cell. \cite{footnoteSCLC1}
We note that these high-quality rubrene single crystals show in-plane field-effect mobilities at the surface of up to 10 cm$^2$/V s. \cite{claudia04}

An optimized sample preparation method was found in a slightly adapted ``flip-crystal'' approach, \cite{Takeya2003} benefiting from the minimized sample handling. The thin crystals (preferred thickness $<$2\,$\mu$m) are placed on glass substrates with 20\,nm Au on 5\,nm Cr electrode stripes, where they stick by electrostatic adhesion. A 20\,nm gold top electrode is then evaporated onto the crystal, which is slightly cooled during the evaporation ($T_{\rm mask}$\,=\,$-8\,^{\circ}$C) to minimize thermal damage. The overlap of the electrodes results in a typical measurement cross section of $A\sim 2\cdot 10^{-5}$\,cm$^2$. Finally, electrical connections to a sample holder were made with silver epoxy and 25-$\mu$m-thick gold wires. The thickness of the crystals was measured by atomic force microscopy, as optical inspection turned out to be unreliable for the ultrathin crystals.

The electrical measurements were performed in a closed-cycle cryostat in inert helium atmosphere and in darkness, covering a maximal temperature range of 30--350\,K. By the use of helium exchange gas, a constant and reliable sample temperature, measured at the back side of the sample holder, could be adjusted with a resistive heater coil (driven by a Lakeshore 331 controller). A Keithley 6517A electrometer was used for the electrical measurements. By a proper shielding, leakage currents well below 1\,pA at 100\,V were achieved. To inject holes from the bottom contact, a negative voltage was applied at the top contact; $V>0$ results in hole injection from the top contact. The superiority of laminated vs evaporated contacts was reported previously. \cite{Zaumseil2003, Boer02}

The typical measurement procedure for TDSCLC was as follows. First, initial tests at room temperature were performed in order to check the reproducibility of repeated measurements of the $j(U)$ characteristic. Thereafter the sample was cooled to typically 100\,K at a rate of 2\,K/min. Prior to the measurement of the current-voltage characteristic (at each chosen temperature), an initial delay of 20\,min ensured thermal equilibrium. The $j(U)$ curves were measured by a stepwise increase of the applied voltage and measuring the quasiequilibrium current. A current measurement delay of 10\,s turned out to be sufficiently long compared to the settling time of the system. Typically 50 points were measured per voltage decade. The maximal voltage was limited such that the current density would not exceed 0.1\,A/cm$^2$ in order to avoid crystal damage. 

Usually $j(U)$ was measured every 10\,K between 100 and 200\,K. At lower temperatures it is difficult to get accurate $j(U)$ curves, because the current increases extremely rapidly with voltage. At higher temperatures, the broadening of the Fermi statistics has an adverse influence on the spectroscopic character of TDSCLC. In particular, shallow states can only be reliably measured at low temperature.

The issue of possible long-term charge trapping as result of a $j(U)$ sweep needs careful attention during the course of a measurement. In most crystals, subsequent sweeps at the same low temperature yield identical curves. This indicates a detrapping of the charge on a time scale faster than that of a sweep ($\sim 10$ min). For a few samples, however, it was necessary to warm the crystal to room temperature after every sweep in order to restore the initial condition. Subsequent $j(U)$ sweeps at the same low temperature then yield the same results. If charge trapping or sample deterioration influence the measurement, the current at a given fixed voltage will not be thermally activated (as opposed to the Arrhenius plot in Fig.~\ref{FigArr}). Therefore, the measurement itself is a valuable self-consistency test.  

\newpage
\section{Experimental results}
\subsection{DOS in rubrene single crystals}

\begin{figure}
\includegraphics[width=8.5cm]{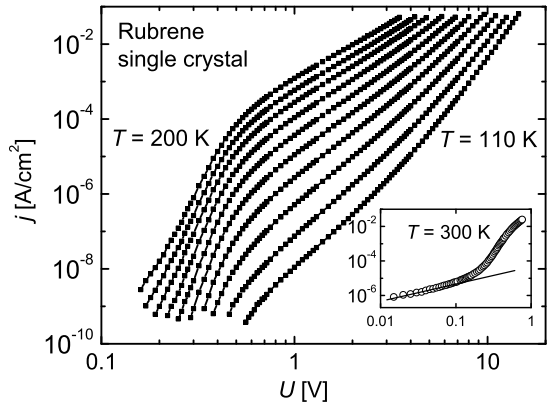}
 \caption{\label{FigTMSCLC}Space-charge-limited-current (SCLC) density vs applied voltage at different temperatures for a rubrene single crystal (Ru65-2, $L=0.6$ $\mu$m). The temperature step is 10 K. The inset shows $j(U)$ at 300 K. The straight line indicates the Ohmic behavior of thermally generated charge carriers.}
\end{figure}

In Fig.~\ref{FigTMSCLC}, the $j(U)$ curves at different temperatures are shown, measured perpendicular to the molecule layers in a rubrene single crystal (Ru65-2, $L=0.6$ \nolinebreak $\mu$m). The ``Ohmic'' region at low voltages is indicated by a straight line (see inset of Fig.~\ref{FigTMSCLC}). The onset of SCLC at $\sim 0.1$\,V indicates that holes are effectively injected from the laminated gold contact. At lower temperatures the number of thermally excited carriers decreases exponentially and the Ohmic region disappears below the sensitivity of the measurement setup ($\sim 10^{-14}$\,A).

\begin{figure}
\includegraphics[width=8.5cm]{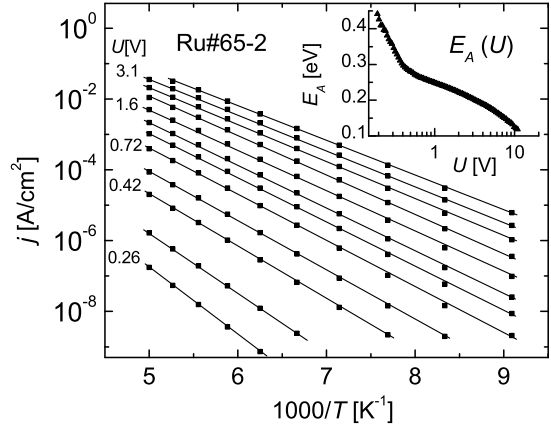}
 \caption{\label{FigArr} Current density for a fixed applied voltage $U$
   vs the inverse temperature for the data plotted in
   Fig. \ref{FigTMSCLC}. The straight lines are the Arrhenius fits
   used to determine the activation energy $E_{\rm A}(U)$. The inset shows
   the resulting $E_{\rm A}(U)$.}
\end{figure}

In Fig.~\ref{FigArr} the Arrhenius plots of the current density are shown for the same data set as in Fig.~\ref{FigTMSCLC}. For clarity we show only data points for selected voltages. The slope yields the activation energy $E_{\rm A}(U)$ for each applied voltage $U$ and we note the high quality of these plots. The resulting $E_{\rm A}(U)$ is given in the inset and is in agreement with recent FET data \cite{Butko2005}. The effective Fermi energy is moved from $\sim 0.45$\,eV at the lowest voltage to $\sim 0.1$\,eV at the highest injection voltage. The distinct change of slope near 0.27\,eV is worth noting, as it obviously reflects a marked increase of the trap density. We note that, among other evidence, the smooth variation of $E_{\rm A}$ at low $U$ indicates space-charge-limited-transport rather than transport limited by contacts. \cite{Nespurek08}

\begin{figure}
\includegraphics[width=8.5cm]{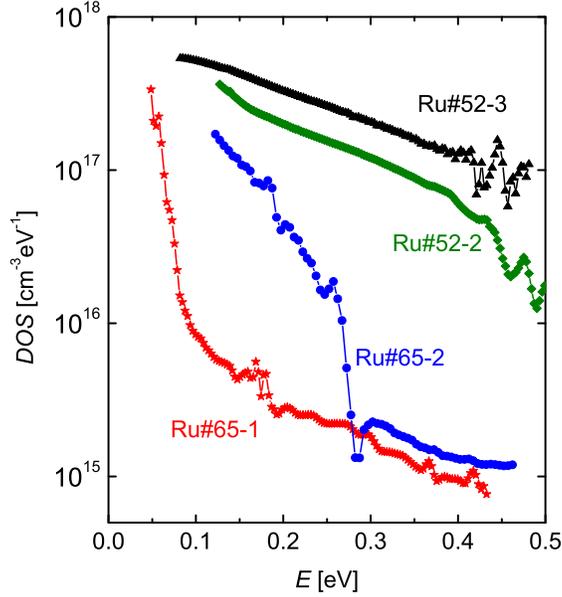}
\caption{\label{Figh}(Color online) Density of states above the valence band for four rubrene single crystals.}
\end{figure}

The analysis described above is used to extract the density of trap states from these TDSCLC measurements. In order to calculate higher-order derivatives of the experimental $j(U)$ and $E_{\rm A}(U)$ curves according to Eq.\,(\ref{dns}), a smoothing spline fit was applied to the measured data, keeping the fit within 1\% of the raw data. The resulting density of trap states after the deconvolution is shown in Fig.~\ref{Figh}. The edge of the valence band is used as the reference level and the positive energy axis points to the center of the band gap. Three main features can be discerned. (1) An exponential increase of the DOS toward the band is observed for all rubrene single crystals. However, the characteristic energy $k_{\rm B}T_{\rm t}$ over which the DOS is reduced by a factor of $e$ varies from sample to sample. The sample Ru52-3 has a broad exponential distribution with $k_{\rm B}T_{\rm t}=210$\,meV and the highest density of trap states. Broad tail states with $k_{\rm B}T_{\rm t}=180$\,meV were recently reported for pentacene single crystals \cite{Lang02}. The sample Ru65-1 with the lowest deep trap density of order $10^{15}$ \nolinebreak cm$^{-1}$\,eV$^{-1}$ in the energy range from 0.45 to 0.1\,eV has a characteristic exponential distribution parameter of $k_{\rm B}T_{\rm t}=180$\,meV. (2) Of particular interest in this sample are the shallow trap states below $\sim 0.1$\,eV with a steeper slope ($k_{\rm B}T_{\rm t}=11$\,meV), reminiscent of band tail states. Due to the large increase of the trap density the quasi Fermi level is pinned close to the band edge, and the amount of injected charge cannot fill these shallow tail states. In this energy range the charge transport is still activated ($E_{\rm A} \sim 0.05$\,eV). (3) The fine structure of the DOS indicates small features for all crystals, due to discrete trap levels in the band gap. The observed fine structure was confirmed using the energy dependence of the statistical shift. \cite{Schauer04}

In comparison, data of organic thin films show a very similar qualitative DOS spectrum, \cite{Tal2005} exhibiting tail states and exponentially increasing deep states, albeit at much higher concentrations than in the purest rubrene single crystals. Again, the FET geometry in Ref.~\onlinecite{Tal2005} might have emphasized defects near the interface, because FET devices made from the same type of high quality rubrene single crystals show approximately three orders of magnitude higher interface trap density than the bulk trap density presented here.  \cite{claudia06}

It is remarkable that the representative selection of rubrene crystals vary in their DOS, although grown under basically identical conditions. \cite{footnoteSCLC2} This variation may originate from individual micro-conditions during and after crystal growth, i.e., different actual growth temperature, growth rate, (thermo)mechanical strain, and different atmospheric conditions in the device fabrication process. On the other hand, measurements of different cross sections on the {\em same\/} crystal result in virtually identical DOS spectra, which is a convincing verification of the data evaluation.

\subsection{Oxygen-induced bulk trap states}

In order to understand the role of bulk traps in organic single crystals and to demonstrate the power of TDSCLC spectroscopy, we investigated the influence of oxygen on the trap density of rubrene single crystals. It is well known that the reaction of rubrene with singlet oxygen $^1$O$_2$ forms the endoperoxide. \cite{Takahashi01} Rubrene itself can act as the sensitizer when the triplet state of rubrene is populated by an intersystem crossing from the singlet state excited with visible light. The energy of this triplet state at 1.2 eV above the ground state is transferred to the molecular oxygen resulting in $^1$O$_2$ which reacts with the rubrene molecule. 

\begin{figure}
\includegraphics[width=8.5cm]{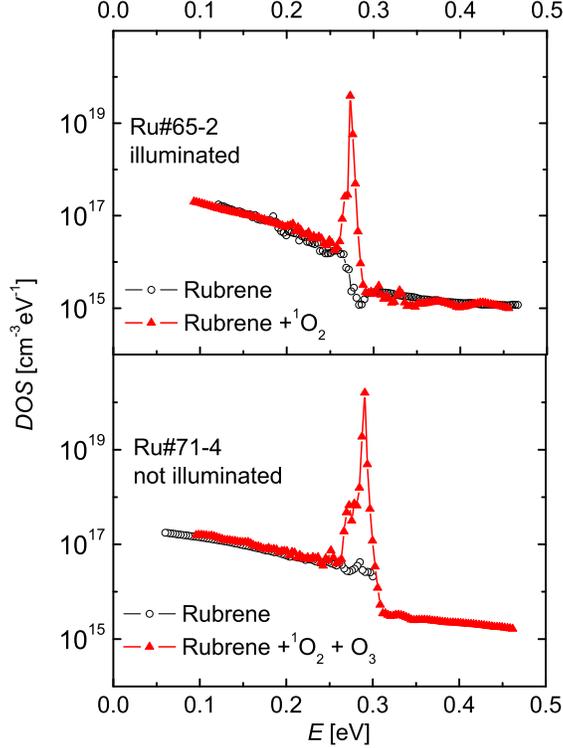}
\caption{\label{FighO2}(Color online) Density of trap states in rubrene single crystals before and after exposure to $^1$O$_2$. The oxygen-induced defect acts as a hole trap at an energy of 0.27\,eV above the valence band edge. Sample Ru65-2 was illuminated in an oxygen atmosphere to form the $^1$O$_2$ directly at the rubrene crystal surface. Sample Ru71-4 was exposed to oxygen excited by uv light.}
\end{figure}

After measuring the full DOS in the as-grown crystals (open symbols in Fig. \ref{FighO2}), we illuminated the sample Ru65-2 with visible light under oxygen atmosphere for four hours. For comparison the sample Ru71-4 was directly exposed for four hours to $^1$O$_2$ by exciting molecular oxygen with uv light in the vicinity of the sample; the sample itself was held in the dark. The exposure of rubrene to $^1$O$_2$ results in a large peak in the density of trap states at 0.27 eV above the valence band (filled symbols in Fig. \ref{FighO2}). The amount of oxygen-induced trap states in sample Ru65-2 is estimated to be $N_{\rm t}^{\rm ox}\approx 2\cdot10^{17}$\,cm$^{-3}$, which corresponds to $\sim 100$ ppm. Because the sample was illuminated in an oxygen atmosphere without removing from the cryostat we can exclude origins other than oxygen for this hole trap. We note that the energetic position of the identified oxygen trap state is in agreement with recently published photoluminescence measurements on oxydized rubrene. \cite{Mitrofanov2006} In sample Ru71-4 the oxygen-induced trap states are within the measurement error at the same energy, but the density of the trap states is slightly higher than for the illuminated sample ($N_{\rm t}^{\rm ox}\approx 3\cdot10^{17}$\,cm$^{-3}$). In addition a shoulder appears in the energetic distribution on the side closer to the valence band. This difference may be due to the formation of O$_3$ under uv light which might also react with the rubrene single crystal. 

The trap level at $E=0.27$\,eV already exists in the pristine rubrene single crystals, with much lower concentrations. This is due to the device fabrication process; the samples are handled in room air under microscope illumination, which results in a similar reaction as for the illuminated sample Ru65-2. The oxygen-related trap level was also observed in other rubrene samples with various concentrations, probably as a result of a different exposure time to air during handling.

\section{Conclusions}

In conclusion, we have implemented temperature-dependent space-charge-limited-current spectroscopy and demonstrate it to be a powerful tool to quantitatively measure the density of bulk trap states with high energy resolution. Applied to high-quality rubrene crystals this method reveals the existence of states within $\sim 0.1$ eV of the band edge, reminiscent of band tails, and a smooth distribution of states deeper in the gap. Discrete peaks are also observed, and through a controlled exposure of the crystals to activated oxygen, a distinct and stable trap level at 0.27\,eV has been created. Applying this approach to other organic semiconductors will be very helpful in the quest to identify intrinsic and extrinsic factors that dominate charge transport in organic semiconductors.

\section*{Acknowledgments}

The authors thank H.-P. Staub and K. Mattenberger for technical assistance with crystal growth and the low-temperature measurements and D. Oberhoff for helpful discussions and support with MATLAB.


\begin{thebibliography}{30}
\expandafter\ifx\csname natexlab\endcsname\relax\def\natexlab#1{#1}\fi
\expandafter\ifx\csname bibnamefont\endcsname\relax
  \def\bibnamefont#1{#1}\fi
\expandafter\ifx\csname bibfnamefont\endcsname\relax
  \def\bibfnamefont#1{#1}\fi
\expandafter\ifx\csname citenamefont\endcsname\relax
  \def\citenamefont#1{#1}\fi
\expandafter\ifx\csname url\endcsname\relax
  \def\url#1{\texttt{#1}}\fi
\expandafter\ifx\csname urlprefix\endcsname\relax\def\urlprefix{URL }\fi
\providecommand{\bibinfo}[2]{#2}
\providecommand{\eprint}[2][]{\url{#2}}

\bibitem[{\citenamefont{Horowitz}(1998)}]{Horowitz1998}
\bibinfo{author}{\bibfnamefont{G.}~\bibnamefont{Horowitz}},
  \bibinfo{journal}{Adv. Mater.} \textbf{\bibinfo{volume}{10}},
  \bibinfo{pages}{365 } (\bibinfo{year}{1998}).

\bibitem[{\citenamefont{Gundlach et~al.}(1997)\citenamefont{Gundlach, Lin,
  Jackson, Nelson, and Schlom}}]{Gundlach02}
\bibinfo{author}{\bibfnamefont{D.~J.} \bibnamefont{Gundlach}},
  \bibinfo{author}{\bibfnamefont{Y.~Y.} \bibnamefont{Lin}},
  \bibinfo{author}{\bibfnamefont{T.~N.} \bibnamefont{Jackson}},
  \bibinfo{author}{\bibfnamefont{S.~F.} \bibnamefont{Nelson}},
  \bibnamefont{and} \bibinfo{author}{\bibfnamefont{D.~G.}
  \bibnamefont{Schlom}}, \bibinfo{journal}{Electron Device Letters, IEEE}
  \textbf{\bibinfo{volume}{18}}, \bibinfo{pages}{87} (\bibinfo{year}{1997}).

\bibitem[{\citenamefont{Dimitrakopoulos and Malenfant}(2002)}]{Dimi01}
\bibinfo{author}{\bibfnamefont{C.~D.} \bibnamefont{Dimitrakopoulos}}
  \bibnamefont{and} \bibinfo{author}{\bibfnamefont{P.~R.~L.}
  \bibnamefont{Malenfant}}, \bibinfo{journal}{Adv. Mater.}
  \textbf{\bibinfo{volume}{14}}, \bibinfo{pages}{99} (\bibinfo{year}{2002}).

\bibitem[{\citenamefont{Kelley et~al.}(2003)\citenamefont{Kelley, Muyres,
  Baude, Smith, and Jones}}]{Kelley2003}
\bibinfo{author}{\bibfnamefont{T.~W.} \bibnamefont{Kelley}},
  \bibinfo{author}{\bibfnamefont{D.~V.} \bibnamefont{Muyres}},
  \bibinfo{author}{\bibfnamefont{P.~F.} \bibnamefont{Baude}},
  \bibinfo{author}{\bibfnamefont{T.~P.} \bibnamefont{Smith}}, \bibnamefont{and}
  \bibinfo{author}{\bibfnamefont{T.~D.} \bibnamefont{Jones}}, in
  \emph{\bibinfo{booktitle}{Mat. Res. Soc. Symp. Proc. Vol. 771}}
  (\bibinfo{year}{2003}), p. \bibinfo{pages}{L6.5.1}.

\bibitem[{\citenamefont{Klauk et~al.}(2002)\citenamefont{Klauk, Halik,
  Zschieschang, Schmid, Radlik, and Weber}}]{Klauk02}
\bibinfo{author}{\bibfnamefont{H.}~\bibnamefont{Klauk}},
  \bibinfo{author}{\bibfnamefont{M.}~\bibnamefont{Halik}},
  \bibinfo{author}{\bibfnamefont{U.}~\bibnamefont{Zschieschang}},
  \bibinfo{author}{\bibfnamefont{G.}~\bibnamefont{Schmid}},
  \bibinfo{author}{\bibfnamefont{W.}~\bibnamefont{Radlik}}, \bibnamefont{and}
  \bibinfo{author}{\bibfnamefont{W.~J.} \bibnamefont{Weber}},
  \bibinfo{journal}{J. Appl.\ Phys.} \textbf{\bibinfo{volume}{92}},
  \bibinfo{pages}{5259} (\bibinfo{year}{2002}).

\bibitem[{\citenamefont{Gershenson et~al.}(2006)\citenamefont{Gershenson,
  Podzorov, and Morpurgo}}]{Gershenson:2006}
\bibinfo{author}{\bibfnamefont{M.~E.} \bibnamefont{Gershenson}},
  \bibinfo{author}{\bibfnamefont{V.}~\bibnamefont{Podzorov}}, \bibnamefont{and}
  \bibinfo{author}{\bibfnamefont{A.~F.} \bibnamefont{Morpurgo}},
  \bibinfo{journal}{Rev. Mod. Phys.} \textbf{\bibinfo{volume}{78}},
  \bibinfo{eid}{973} (\bibinfo{year}{2006}).

\bibitem[{\citenamefont{Warta and Karl}(1985)}]{Warta1985}
\bibinfo{author}{\bibfnamefont{W.}~\bibnamefont{Warta}} \bibnamefont{and}
  \bibinfo{author}{\bibfnamefont{N.}~\bibnamefont{Karl}},
  \bibinfo{journal}{Phys. Rev. B} \textbf{\bibinfo{volume}{32}},
  \bibinfo{pages}{1172} (\bibinfo{year}{1985}).

\bibitem[{\citenamefont{Lang et~al.}(2004{\natexlab{a}})\citenamefont{Lang,
  Chi, Siegrist, Sergent, and Ramirez}}]{Lang01}
\bibinfo{author}{\bibfnamefont{D.~V.} \bibnamefont{Lang}},
  \bibinfo{author}{\bibfnamefont{X.}~\bibnamefont{Chi}},
  \bibinfo{author}{\bibfnamefont{T.}~\bibnamefont{Siegrist}},
  \bibinfo{author}{\bibfnamefont{A.~M.} \bibnamefont{Sergent}},
  \bibnamefont{and} \bibinfo{author}{\bibfnamefont{A.~P.}
  \bibnamefont{Ramirez}}, \bibinfo{journal}{Phys. Rev. Lett.}
  \textbf{\bibinfo{volume}{93}}, \bibinfo{pages}{076601}
  (\bibinfo{year}{2004}{\natexlab{a}}).

\bibitem[{\citenamefont{Mitrofanov et~al.}(2006)\citenamefont{Mitrofanov, Lang,
  Kloc, Wikberg, Siegrist, So, Sergent, and Ramirez}}]{Mitrofanov2006}
\bibinfo{author}{\bibfnamefont{O.}~\bibnamefont{Mitrofanov}},
  \bibinfo{author}{\bibfnamefont{D.~V.} \bibnamefont{Lang}},
  \bibinfo{author}{\bibfnamefont{C.}~\bibnamefont{Kloc}},
  \bibinfo{author}{\bibfnamefont{J.~M.} \bibnamefont{Wikberg}},
  \bibinfo{author}{\bibfnamefont{T.}~\bibnamefont{Siegrist}},
  \bibinfo{author}{\bibfnamefont{W.-Y.} \bibnamefont{So}},
  \bibinfo{author}{\bibfnamefont{M.~A.} \bibnamefont{Sergent}},
  \bibnamefont{and} \bibinfo{author}{\bibfnamefont{A.~P.}
  \bibnamefont{Ramirez}}, \bibinfo{journal}{Phys. Rev. Lett.}
  \textbf{\bibinfo{volume}{97}}, \bibinfo{pages}{166601}
  (\bibinfo{year}{2006}).

\bibitem[{\citenamefont{Tal et~al.}(2005)\citenamefont{Tal, Rosenwaks,
  Preezant, Tessler, Chan, and Kahn}}]{Tal2005}
\bibinfo{author}{\bibfnamefont{O.}~\bibnamefont{Tal}},
  \bibinfo{author}{\bibfnamefont{Y.}~\bibnamefont{Rosenwaks}},
  \bibinfo{author}{\bibfnamefont{Y.}~\bibnamefont{Preezant}},
  \bibinfo{author}{\bibfnamefont{N.}~\bibnamefont{Tessler}},
  \bibinfo{author}{\bibfnamefont{C.~K.} \bibnamefont{Chan}}, \bibnamefont{and}
  \bibinfo{author}{\bibfnamefont{A.}~\bibnamefont{Kahn}},
  \bibinfo{journal}{Phys. Rev. Lett.} \textbf{\bibinfo{volume}{95}},
  \bibinfo{eid}{256405} (\bibinfo{year}{2005}).

\bibitem[{\citenamefont{Schauer et~al.}(1996)\citenamefont{Schauer,
  Ne\^sp$\mathring{\textnormal{u}}$rek, and Valeri\'an}}]{Schauer04}
\bibinfo{author}{\bibfnamefont{F.}~\bibnamefont{Schauer}},
  \bibinfo{author}{\bibfnamefont{S.}~\bibnamefont{Ne\^sp$\mathring{\textnormal%
{u}}$rek}}, \bibnamefont{and}
  \bibinfo{author}{\bibfnamefont{H.}~\bibnamefont{Valeri\'an}},
  \bibinfo{journal}{J. Appl. Phys.} \textbf{\bibinfo{volume}{80}},
  \bibinfo{pages}{880} (\bibinfo{year}{1996}).

\bibitem[{\citenamefont{Arkhipov et~al.}(2001)\citenamefont{Arkhipov, Heremans,
  Emelianova, and Adriaenssens}}]{Arkhipov01}
\bibinfo{author}{\bibfnamefont{V.~I.} \bibnamefont{Arkhipov}},
  \bibinfo{author}{\bibfnamefont{P.}~\bibnamefont{Heremans}},
  \bibinfo{author}{\bibfnamefont{E.~V.} \bibnamefont{Emelianova}},
  \bibnamefont{and} \bibinfo{author}{\bibfnamefont{G.~J.}
  \bibnamefont{Adriaenssens}}, \bibinfo{journal}{Appl. Phys. Lett}
  \textbf{\bibinfo{volume}{79}}, \bibinfo{pages}{4154} (\bibinfo{year}{2001}).

\bibitem[{\citenamefont{Schauer et~al.}(1997)\citenamefont{Schauer, Novotny,
  and Ne\^sp$\mathring{\textnormal{u}}$rek}}]{Schauer03}
\bibinfo{author}{\bibfnamefont{F.}~\bibnamefont{Schauer}},
  \bibinfo{author}{\bibfnamefont{R.}~\bibnamefont{Novotny}}, \bibnamefont{and}
  \bibinfo{author}{\bibfnamefont{S.}~\bibnamefont{Ne\^sp$\mathring{\textnormal%
{u}}$rek}}, \bibinfo{journal}{J. Appl. Phys.} \textbf{\bibinfo{volume}{81}},
  \bibinfo{pages}{1244} (\bibinfo{year}{1997}).

\bibitem[{\citenamefont{Deutsch and Beniaminy}(1982)}]{Deutsch01}
\bibinfo{author}{\bibfnamefont{M.}~\bibnamefont{Deutsch}} \bibnamefont{and}
  \bibinfo{author}{\bibfnamefont{I.}~\bibnamefont{Beniaminy}},
  \bibinfo{journal}{Rev. Sci. Instrum.} \textbf{\bibinfo{volume}{53}},
  \bibinfo{pages}{90} (\bibinfo{year}{1982}).

\bibitem[{\citenamefont{Laudise et~al.}(1998)\citenamefont{Laudise, Kloc,
  Simpkins, and Siegrist}}]{Laudise1998}
\bibinfo{author}{\bibfnamefont{R.~A.} \bibnamefont{Laudise}},
  \bibinfo{author}{\bibfnamefont{C.}~\bibnamefont{Kloc}},
  \bibinfo{author}{\bibfnamefont{P.~G.} \bibnamefont{Simpkins}},
  \bibnamefont{and} \bibinfo{author}{\bibfnamefont{T.}~\bibnamefont{Siegrist}},
  \bibinfo{journal}{J. Cryst. Growth} \textbf{\bibinfo{volume}{187}},
  \bibinfo{pages}{449} (\bibinfo{year}{1998}).

\bibitem[{foo()}]{footnoteSCLC1}
\bibinfo{note}{The direction perpendicular to the surface is parallel to the
  crystallographic $a$- \cite{Jurchescu2006a, Chapman2006} or $c$-axis
  \cite{Sundar2004}, depending on the space group setting.}

\bibitem[{\citenamefont{Jurchescu et~al.}(2006)\citenamefont{Jurchescu,
  Meetsma, and Palstra}}]{Jurchescu2006a}
\bibinfo{author}{\bibfnamefont{O.~D.} \bibnamefont{Jurchescu}},
  \bibinfo{author}{\bibfnamefont{A.}~\bibnamefont{Meetsma}}, \bibnamefont{and}
  \bibinfo{author}{\bibfnamefont{T.~T.~M.} \bibnamefont{Palstra}},
  \bibinfo{journal}{Acta Crystallogr. B} \textbf{\bibinfo{volume}{62}},
  \bibinfo{pages}{330} (\bibinfo{year}{2006}).

\bibitem[{\citenamefont{Chapman et~al.}(2006)\citenamefont{Chapman, Checco,
  Pindak, Siegrist, and Kloc}}]{Chapman2006}
\bibinfo{author}{\bibfnamefont{B.~D.} \bibnamefont{Chapman}},
  \bibinfo{author}{\bibfnamefont{A.}~\bibnamefont{Checco}},
  \bibinfo{author}{\bibfnamefont{R.}~\bibnamefont{Pindak}},
  \bibinfo{author}{\bibfnamefont{T.}~\bibnamefont{Siegrist}}, \bibnamefont{and}
  \bibinfo{author}{\bibfnamefont{C.}~\bibnamefont{Kloc}}, \bibinfo{journal}{J.
  Cryst. Growth} \textbf{\bibinfo{volume}{290}}, \bibinfo{pages}{479}
  (\bibinfo{year}{2006}).

\bibitem[{\citenamefont{Sundar et~al.}(2004)\citenamefont{Sundar, Zaumseil,
  Podzorov, Menard, Willett, Someya, Gershenson, and Rogers}}]{Sundar2004}
\bibinfo{author}{\bibfnamefont{V.~C.} \bibnamefont{Sundar}},
  \bibinfo{author}{\bibfnamefont{J.}~\bibnamefont{Zaumseil}},
  \bibinfo{author}{\bibfnamefont{V.}~\bibnamefont{Podzorov}},
  \bibinfo{author}{\bibfnamefont{E.}~\bibnamefont{Menard}},
  \bibinfo{author}{\bibfnamefont{R.~L.} \bibnamefont{Willett}},
  \bibinfo{author}{\bibfnamefont{T.}~\bibnamefont{Someya}},
  \bibinfo{author}{\bibfnamefont{M.~E.} \bibnamefont{Gershenson}},
  \bibnamefont{and} \bibinfo{author}{\bibfnamefont{J.~A.}
  \bibnamefont{Rogers}}, \bibinfo{journal}{Science}
  \textbf{\bibinfo{volume}{303}}, \bibinfo{pages}{1644}
  (\bibinfo{year}{2004}).

\bibitem[{\citenamefont{Goldmann et~al.}(2004)\citenamefont{Goldmann, Haas,
  Krellner, Pernstich, Gundlach, and Batlogg}}]{claudia04}
\bibinfo{author}{\bibfnamefont{C.}~\bibnamefont{Goldmann}},
  \bibinfo{author}{\bibfnamefont{S.}~\bibnamefont{Haas}},
  \bibinfo{author}{\bibfnamefont{C.}~\bibnamefont{Krellner}},
  \bibinfo{author}{\bibfnamefont{K.~P.} \bibnamefont{Pernstich}},
  \bibinfo{author}{\bibfnamefont{D.~J.} \bibnamefont{Gundlach}},
  \bibnamefont{and} \bibinfo{author}{\bibfnamefont{B.}~\bibnamefont{Batlogg}},
  \bibinfo{journal}{J. Appl. Phys.} \textbf{\bibinfo{volume}{96}},
  \bibinfo{pages}{2080} (\bibinfo{year}{2004}).

\bibitem[{\citenamefont{Takeya et~al.}(2003)\citenamefont{Takeya, Goldmann,
  Haas, Pernstich, Ketterer, and Batlogg}}]{Takeya2003}
\bibinfo{author}{\bibfnamefont{J.}~\bibnamefont{Takeya}},
  \bibinfo{author}{\bibfnamefont{C.}~\bibnamefont{Goldmann}},
  \bibinfo{author}{\bibfnamefont{S.}~\bibnamefont{Haas}},
  \bibinfo{author}{\bibfnamefont{K.~P.} \bibnamefont{Pernstich}},
  \bibinfo{author}{\bibfnamefont{B.}~\bibnamefont{Ketterer}}, \bibnamefont{and}
  \bibinfo{author}{\bibfnamefont{B.}~\bibnamefont{Batlogg}},
  \bibinfo{journal}{J. Appl. Phys.} \textbf{\bibinfo{volume}{94}},
  \bibinfo{pages}{5800} (\bibinfo{year}{2003}).

\bibitem[{\citenamefont{Zaumseil et~al.}(2003)\citenamefont{Zaumseil, Baldwin,
  and Rogers}}]{Zaumseil2003}
\bibinfo{author}{\bibfnamefont{J.}~\bibnamefont{Zaumseil}},
  \bibinfo{author}{\bibfnamefont{K.~W.} \bibnamefont{Baldwin}},
  \bibnamefont{and} \bibinfo{author}{\bibfnamefont{J.~A.}
  \bibnamefont{Rogers}}, \bibinfo{journal}{J. Appl. Phys.}
  \textbf{\bibinfo{volume}{93}}, \bibinfo{pages}{6117}
  (\bibinfo{year}{2003}).

\bibitem[{\citenamefont{de~Boer et~al.}(2004)\citenamefont{de~Boer, Jochemsen,
  Klapwijk, Morpurgo, Niemax, Tripathi, and Pflaum}}]{Boer02}
\bibinfo{author}{\bibfnamefont{R.~W.~I.} \bibnamefont{de~Boer}},
  \bibinfo{author}{\bibfnamefont{M.}~\bibnamefont{Jochemsen}},
  \bibinfo{author}{\bibfnamefont{T.~M.} \bibnamefont{Klapwijk}},
  \bibinfo{author}{\bibfnamefont{A.~F.} \bibnamefont{Morpurgo}},
  \bibinfo{author}{\bibfnamefont{J.}~\bibnamefont{Niemax}},
  \bibinfo{author}{\bibfnamefont{A.~K.} \bibnamefont{Tripathi}},
  \bibnamefont{and} \bibinfo{author}{\bibfnamefont{J.}~\bibnamefont{Pflaum}},
  \bibinfo{journal}{J. Appl. Phys.} \textbf{\bibinfo{volume}{95}},
  \bibinfo{pages}{1196} (\bibinfo{year}{2004}).
  
  
\bibitem[{\citenamefont{Butko et~al.}(2005)\citenamefont{Butko, Lashley, and Ramirez}}]{Butko2005} 
  \bibinfo{author}{\bibfnamefont{V.~Y.}~\bibnamefont{Butko}},
  \bibinfo{author}{\bibfnamefont{J.~C.} \bibnamefont{Lashley}},
  \bibnamefont{and} \bibinfo{author}{\bibfnamefont{A.~P.}
  \bibnamefont{Ramirez}}, \bibinfo{journal}{Phys. Rev. B} \textbf{\bibinfo{volume}{72}},
  \bibinfo{pages}{081312R} (\bibinfo{year}{2005}).


\bibitem[{\citenamefont{Ne\^sp$\mathring{\textnormal{u}}$rek
  et~al.}(1984)\citenamefont{Ne\^sp$\mathring{\textnormal{u}}$rek, Zme\^skal,
  and Schauer}}]{Nespurek08}
\bibinfo{author}{\bibfnamefont{S.}~\bibnamefont{Ne\^sp$\mathring{\textnormal{u%
}}$rek}}, \bibinfo{author}{\bibfnamefont{O.}~\bibnamefont{Zme\^skal}},
  \bibnamefont{and} \bibinfo{author}{\bibfnamefont{F.}~\bibnamefont{Schauer}},
  \bibinfo{journal}{Phys. Status Solidi A} \textbf{\bibinfo{volume}{85}},
  \bibinfo{pages}{619} (\bibinfo{year}{1984}).

\bibitem[{\citenamefont{Lang et~al.}(2004{\natexlab{b}})\citenamefont{Lang,
  Chi, Siegrist, Sergent, and Ramirez}}]{Lang02}
\bibinfo{author}{\bibfnamefont{D.~V.} \bibnamefont{Lang}},
  \bibinfo{author}{\bibfnamefont{X.}~\bibnamefont{Chi}},
  \bibinfo{author}{\bibfnamefont{T.}~\bibnamefont{Siegrist}},
  \bibinfo{author}{\bibfnamefont{A.~M.}~\bibnamefont{Sergent}}, \bibnamefont{and}
  \bibinfo{author}{\bibfnamefont{A.~P.} \bibnamefont{Ramirez}},
  \bibinfo{journal}{Phys. Rev. Lett.} \textbf{\bibinfo{volume}{93}},
  \bibinfo{pages}{086802} (\bibinfo{year}{2004}{\natexlab{b}}).

\bibitem[{\citenamefont{Goldmann et~al.}(2006)\citenamefont{Goldmann, Krellner, Pernstich, Haas, Gundlach, and Batlogg}}]{claudia06}
\bibinfo{author}{\bibfnamefont{C.}~\bibnamefont{Goldmann}},
  \bibinfo{author}{\bibfnamefont{C.}~\bibnamefont{Krellner}},
  \bibinfo{author}{\bibfnamefont{K.~P.} \bibnamefont{Pernstich}},
  \bibinfo{author}{\bibfnamefont{S.}~\bibnamefont{Haas}},  
  \bibinfo{author}{\bibfnamefont{D.~J.} \bibnamefont{Gundlach}},
  \bibnamefont{and} \bibinfo{author}{\bibfnamefont{B.}~\bibnamefont{Batlogg}},
  \bibinfo{journal}{J. Appl. Phys.} \textbf{\bibinfo{volume}{99}},
  \bibinfo{pages}{034507} (\bibinfo{year}{2006}).
  
\bibitem[{foo()}]{footnoteSCLC2}
\bibinfo{note}{The samples Ru65-1 and Ru65-2 are from a different growth batch than the samples Ru52-2 and Ru52-3, but with identical growth parameters.}

\bibitem[{\citenamefont{Takahashi et~al.}(1979)\citenamefont{Takahashi, Harada,
  Sato, Seki, Inokuchi, and Fujisawa}}]{Takahashi01}
\bibinfo{author}{\bibfnamefont{T.}~\bibnamefont{Takahashi}},
  \bibinfo{author}{\bibfnamefont{Y.}~\bibnamefont{Harada}},
  \bibinfo{author}{\bibfnamefont{N.}~\bibnamefont{Sato}},
  \bibinfo{author}{\bibfnamefont{K.}~\bibnamefont{Seki}},
  \bibinfo{author}{\bibfnamefont{H.}~\bibnamefont{Inokuchi}}, \bibnamefont{and}
  \bibinfo{author}{\bibfnamefont{S.}~\bibnamefont{Fujisawa}},
  \bibinfo{journal}{Bull. Chem. Soc. Jpn.} \textbf{\bibinfo{volume}{52}},
  \bibinfo{pages}{380} (\bibinfo{year}{1979}).
  


\end{thebibliography}

\end{document}